\documentclass{ws-procs9x6}

\begin{document}

\title{THE FIRST HALOS\footnote{Talk given at IDM 2006, Rhodes, Greece}}
\author{DOMINIK J. SCHWARZ}

\address{Fakult\"at f\"ur Physik, Universit\"at Bielefeld, Postfach 100131, 33501 Bielefeld, Germany\\ 
E-mail:  dschwarz at physik dot uni-bielefeld dot de}

\begin{abstract}
The size and time of formation of the first gravitationally bound objects in the Universe is set by the 
microphysical properties of the dark matter. It is argued that observations seem to favour cold and 
thermal candidates for the main contribution to the dark matter. For that type of dark matter, the size and 
time of formation of the first halos is determined by the elastic cross sections and mass of the CDM 
particles. Consequently, the astrophysics of CDM might allow us to measure some of the fundamental 
parameters of CDM particles. Essential for observations is the survival rate 
and spatial distribution of the very first objetcs, which are currently under 
debate. 
\end{abstract}

\keywords{dark matter; structure formation/}

\bodymatter 

\section{Motivation}

There are at least four motivations to study the issue of the smallest structures and first objects 
in the Universe. Firstly, a scale invariant power spectrum, without cut-off at some small scale, would 
lead to an infinite amount of energy stored in acoustic waves (density fluctuations) in the Universe and 
would be inconsistent. Thus a cut-off at some scale is necessary and it is interesting to know where it is.
Secondly, we lack a fundamental understanding of dark matter. All dark matter candidates give rise to 
the same structure at the largest scales, but differ in their predictions at smaller scales. E.g. light 
neutrinos and weakly interacting massive particles (WIMPs) lead to different structures at large 
subhorizon scales and can be discriminated by present large scale structure observations. Similarly, 
various WIMP candidates  give rise to different smallest scale structure. Thus we hope that the 
astrophysics of dark matter 
might allow us one day to exclude some of the many dark matter candidates. A third motivation are
direct 
and indirect search experiments. Although experimentalists and observers typically deliver limits on 
some cross section as a function of dark matter mass, they actually constrain the rate. However,  the 
rates depend on some cross section times the local dark matter density. The local dark matter density 
might be significantly different from the matter density in the isothermal halo model. It is thus important 
to understand the small scale distribution of dark matter down to the scale of dark matter (also the 
velocity distribution is of concern, but not discussed in this contribution). Finally, a complete 
understanding of hierarchical structure formation must include the study of the very first objects in the 
hierarchy. 
 
\section{Primordial power spectrum}

We can only make sensible statements on  the power of modes that are generated by quantum 
fluctuations during cosmological inflation. Larger modes are certainly not observable today and we 
have no theory for that regime. The smallest primordial modes cross the horizon close to the end of 
inflation. The energy scale at the end of inflation lies between the Planck 
scale and nucleosythesis, which means that the smallest primordial scales are somewhere between $1$ 
mm and $1$ pc, measured in comoving length scales. If we argue that the matter-antimatter asymmetry 
in the Universe cannot be generated below the electroweak scale, the energy scale at the end of 
inflation should be at least $1$ TeV. In that case the maximal mode number is $k_{\rm max}  > 10^6/
\mbox{pc} \approx 1/(0.5 \mbox{au})$. For length scales smaller than the Hubble scale at the end of inflation, 
no squeezing of quantum fluctuations occurs and thus it seems that those modes will never become 
"classical" density inhomogeneities. For our discussion it is enough to realise, that the scenario of 
cosmological inflation, together with a successful mechanism of baryogenesis, allows us to speak about 
the primordial power spectrum down to at least $10^{12}/\mbox{Mpc}$ or, expressed in cold matter 
mass, $10^{-24} M_\odot$.  

\section{Classification of dark matter candidates}

There are basically two important criteria according to which dark matter candidates can be classified. 
The first one introduces the notion of HOT and COLD dark matter, but has nothing to do with temperature. Hot 
dark matter obeys a relativistic equation of state ($p \sim \epsilon$) at the time of matter-radiation 
equality,  whereas cold dark matter is non-relativistic at that moment of time ($p \ll \epsilon$). The matter-
radiation equality is the relevant moment of time, because growth of structure happens in the matter 
dominated epoch only (decoupled dark matter inhomogeneities grow only logarithmically in the 
radiation dominated epoch). 

Whether a dark matter candidate was ever in thermal equilibrium with the hot radiation 
fluid provides a second criterion. If so, it is called THERMAL, if not NON-THERMAL. Some examples of
all four categories are given 
in table \ref{tab1}. 
 
\begin{table}[t]
\tbl{\label{tab1}Classification of dark matter candidates} 
{\begin{tabular}{|l|c|c|}
\hline
        & HOT (relativistic) & COLD (non-relativistic) \\
\hline
THERMAL & light $\nu$s, \dots & WIMPs (heavy $\nu$, LSP, \dots),  \\
                      &                                 & \dots \\
\hline 
NON-THERMAL & string gas, \dots & misalignment axions,  \\
            &                   & primordial black holes, \dots \\
\hline 
\end{tabular}}
\end{table}

\section{What is the dominant dark matter?}

We may ask, given that classification, is any one favoured by observation? A key here is that 
thermal candidates must have isentropic (often called adiabatic) initial conditions, whereas the non-
thermal ones can have more general initial conditions (isocurvature/entropy perturbations). From WMAP 
and large scale structure data, it follows that the isentropic mode must be the dominant one, a 
contribution of order $10\%$ of isocurvature modes is not excluded (see e.g.~\cite{Trotta}) . This 
suggests that thermal dark matter candidates are compatible with observations, but non-thermal candidates must provide a 
mechanism that surpresses isocurvature modes. An especially interesting 
case in this aspect is the axion, which seems on the first sight to be disfavoured by the 
data, as it allows isocurvature perturbations. However, as the onset of axion mass is triggered by the temperature of the coloured degrees of freedom 
in the hot plasma, the axion density perturbations are also largely isentropic, if inflation happens 
after Peccei-Quinn symmetry breaking. In the case of PQ symmetry breaking after inflation, the axion 
comes with isocurvature perturbations (see e.g.\cite{Sikivie}). 
To summarise, the data do not exclude non-thermal candidates in general, but they require that their isocurvature power is suppressed to the few \% level.

What remains is the question hot versus cold dark matter. The current limits on light neutrinos can be 
taken representative for any hot and thermal component. They would lead to a different matter-radiation 
equality and to a damping of smaller structures due to neutrino free streeming. Fitting WMAP data and 
large scale structrure data indicates that not more than $10\%$  of the dark matter can be hot (see e.g. 
\cite{Tegmark}).
We thus conclude that observations are most naturally explained by candidates that are classified as 
cold and thermal dark matter. There are plenty of candidates in that category: the lightest 
supersymmetric particle (in many scenarios the neutralino), the lightest Kaluza-Klein  particle,  a heavy 
fourth generation neutrino (mass of order TeV), any other WIMP, etc.  
 
\section{Thermal and non-relativistic dark matter}

In the folloing, we focus our attention on this class of candidates. 

\subsection{Kinetic decoupling}

In contrast to relativistic thermal dark matter, chemical decoupling (freeze-out) and the kinetic 
decoupling happens at different  moments of time \cite{SSW2,HSS,Berezinsky}. For WIMPs with 
mass $m$, the freeze out happens at 
$T_{\rm cd} \sim m/25$, whereas kinetic decoupling happens much later; in order
to keep thermal equilibrium only elastic scattering is necessary. (A detailed discussion of the relevant elastic rates for neutralinos is given in \cite{Chen,HSS}.) The decoupling time can be estimated 
with help of the relaxation time $\tau_{\rm relax}$. In each collision the typical change of momentum is 
\[
\frac{\Delta p}{p} \sim  \sqrt{\frac{T}{m}}.
\]
The CDM particle makes a random walk in momentum space, $(\Delta p/p)_N \sim \sqrt{N} \Delta p/p$,
and thus has to scatter $N \sim m/T$ times to significantly change to the WIMP's momentum. When 
the relaxation time $
\tau_{\rm relax} \sim N \tau_{\rm coll} $
reaches the Hubble time $t_{\rm H}$ the CDM decouples. For a WIMP with elastic cross sections 
\[
\sigma_{\rm el} \sim \frac{m^2}{m_Z^4} \left(\frac Tm\right)^{1 + l},
\]
where $l = 0$ or $1$ for domination of s- or p-wave scattering, we find that kinetic decoupling happens 
at
\[
T_{\rm kd} \sim \left(\frac{m_Z^4 m^l}{m_{\rm Pl}}\right)^{\frac{1}{3+l}}, 
\]
ranging from a few MeV to a few 10 MeV \cite{GHS1,GHS2}. 
Profumo et al. \cite{Profumo} performed a more detailed
calculation for specific models of supersymmetry and found that the decoupling temperture might be as 
large as a few GeV. They also invesitgated models with universal extra dimensions, which are in 
concordance with our naive estimates.   

More recently Bertschinger \cite{Bertschinger} improved our estimate of the decoupling temperature by 
solving the appropriate Fokker-Planck equation. His results agree with our estimates at $\sim  20\%$.   

\subsection{Damping and growth of inhomogeneities}

During and after kinetic decoupling  the primordial inhomogeneities (from cosmological inflation) are 
washed out on the smallest scales. Before decoupling the CDM starts to deviate from the prefect fluid 
behaviour of the radiation fluid and this is very well described by an imperfect fluid initially. Hofmann et 
al.\cite{HSS} calculated  the viscosities and used the linearised 
Navier-Stokes equations to obtain a first estimate of this collisional damping epoch. 

After decoupling the CDM particles continue to damp the smallest inhomogeneities due to free 
streaming\cite{HSS,Berezinsky,Boehm,Loeb}.  The comoving free steaming length $l_{\rm fs}$ is a function of time, but becomes 
approximately constant as $z \ll z_{\rm eq}$.  It defines the damping 
scale, which, expressed as the mass of CDM in a homogeneous sphere of radius $l_{\rm fs}/2$, is 
estimated as 
\[
M_{\rm fs}  \sim 10^{-6} M_\odot \left(\frac{(1+ \ln(T_{\rm kd}/30 \mbox{\ MeV})/19.2)^2}{(m/(100 \mbox{\ GeV})(T_{\rm kd}/ 30 \mbox{\ MeV})}
\right)^{3/2} .
\]
Here we assumed a matter density of $\omega_{\rm m} = 0.14$. 
Collisional damping and free streaming have been incorporated by Green et al. \cite{GHS1,GHS2}  to 
estimate a typical cut-off scale in the CDM power spectrum. It turns out that for the generic WIMP 
the free streaming is the dominant contribution to damping and thus sets the scale for the smallest 
structures.  

Loeb and Zaldarriaga \cite{Loeb} pointed out correctly, that the acoustic oscillations and the 
drag of the radiation fluid cannot be neglected at decoupling itself and they showed numerically that this 
has the effect to give rise to even more damping, however the order of magnitue is correctly estimated by 
free streaming only. Most recently Bertschinger \cite{Bertschinger} showed that the ad hoc treatment of 
the radiation drag in reference \cite{Loeb}  can be improved with help of the appropriate the Fokker-
Planck equation.  These studies lead consistently to the conclusion that the order of magnitude of the 
cut-off in wave number is provided by the free streaming calculation. To fix the order of 
magnitude of the mass scale a more detailed calculation is needed, as a factor of $2$ in wave number is 
a factor of $8$ in mass!  Close to the cut-off, the CDM power spectrum inherits the peaks and dips from  the acoustic oscillations in the radiation fluid. The typical wavelength of these oscillations is given 
by the sound horizon at kinetic decoupling, in terms of CDM mass 
\[ 
M_{\rm osc} \sim 10^{-5} M_\odot \left( \frac{30 \mbox{\ MeV}}{T_{\rm kd}}\right)^3.
\] 

The damping of inhomogeneities must be convoluted with the primordial spectrum to provide the 
cut-off scale and the scale of the maximal primordial power \cite{GHS2}. The CDM becomes fully decoupled  only after kinetic decoupling and inhomogeneities in the CDM start to grow logarithmically. This is most easily 
seen from the evolution equation on subhorizon scales
\[
a^2 \frac{{\rm d}^2 \Delta_{\rm cdm}}{{\rm d} a^2} + \frac 3 2 \left( 1 - \frac p \epsilon \right) a 
\frac{{\rm d} \Delta_{\rm cdm}}{{\rm d} a} - \frac 3 2 \frac{\epsilon_{\rm cdm}}{\epsilon} \Delta_{\rm cdm} = 
0,
\]
where $\Delta_{\rm cdm}$ is the CDM density contrast, $a$ the scale factor, and $\epsilon$ and $p$ are 
energy density and pressure, respectively. During the radiation dominated epoch 
$\epsilon_{\rm cdm} \ll \epsilon$ and $p \approx \epsilon/3$, which leads to $\Delta_{\rm cdm} = 
A + B \ln (a/a_{\rm kd})$, where A and B are constants to be fixed by the initial conditions. It is also 
possible to take the effects of baryons into account analytically. After CDM kinetic decoupling they are 
still tightly coupled to the radiation fluid and suppress the growth of structure until they decouple at 
around $z_{\rm b} \sim 150$. 

Figure 1 shows the influence of the WIMP microphysics,
whereas figure 2 shows how the maximum depends on the details of the primordial power spectrum. 
\begin{figure}[t]
\psfig{file=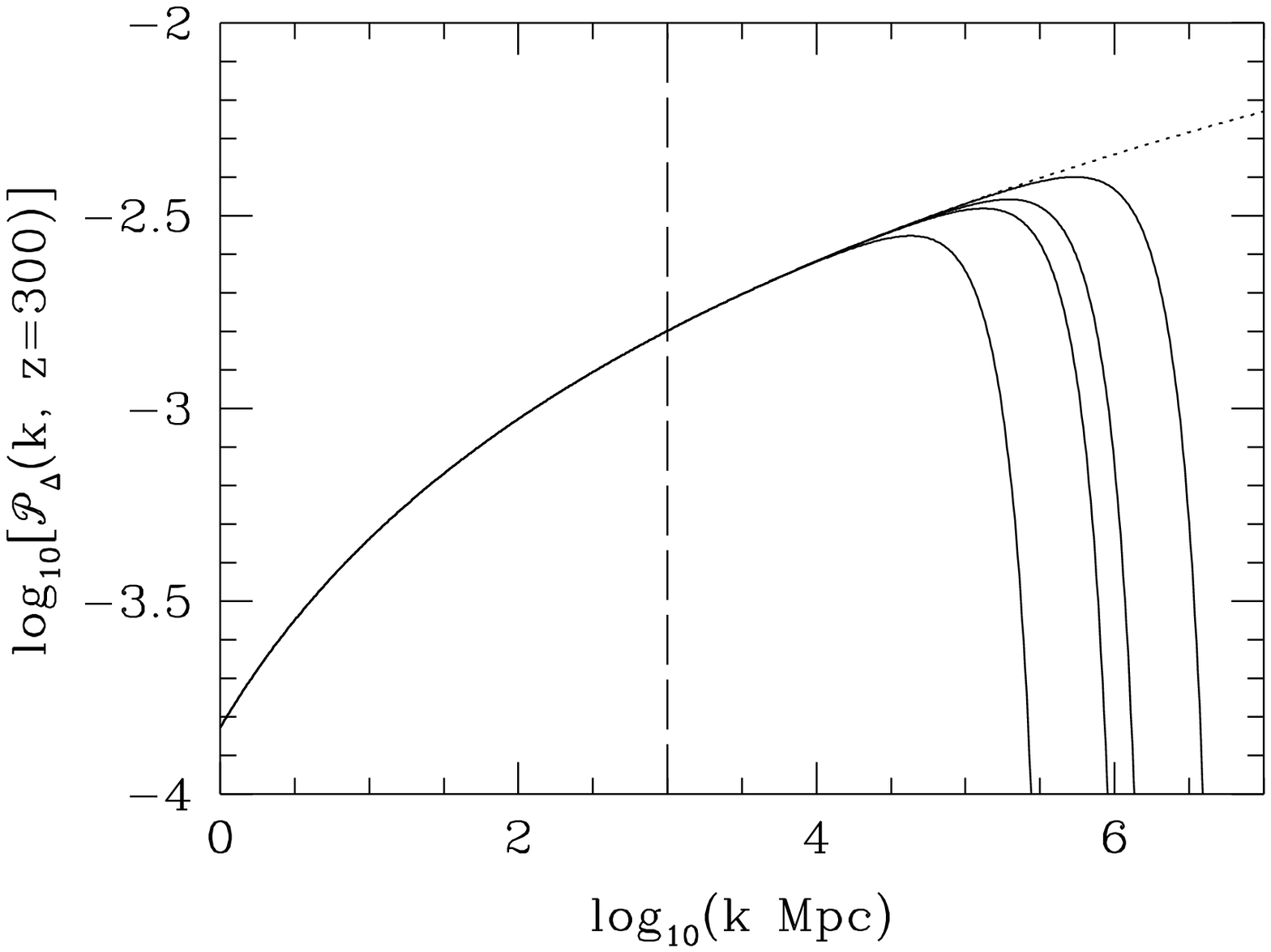, width=5 cm}\hfill
\psfig{file=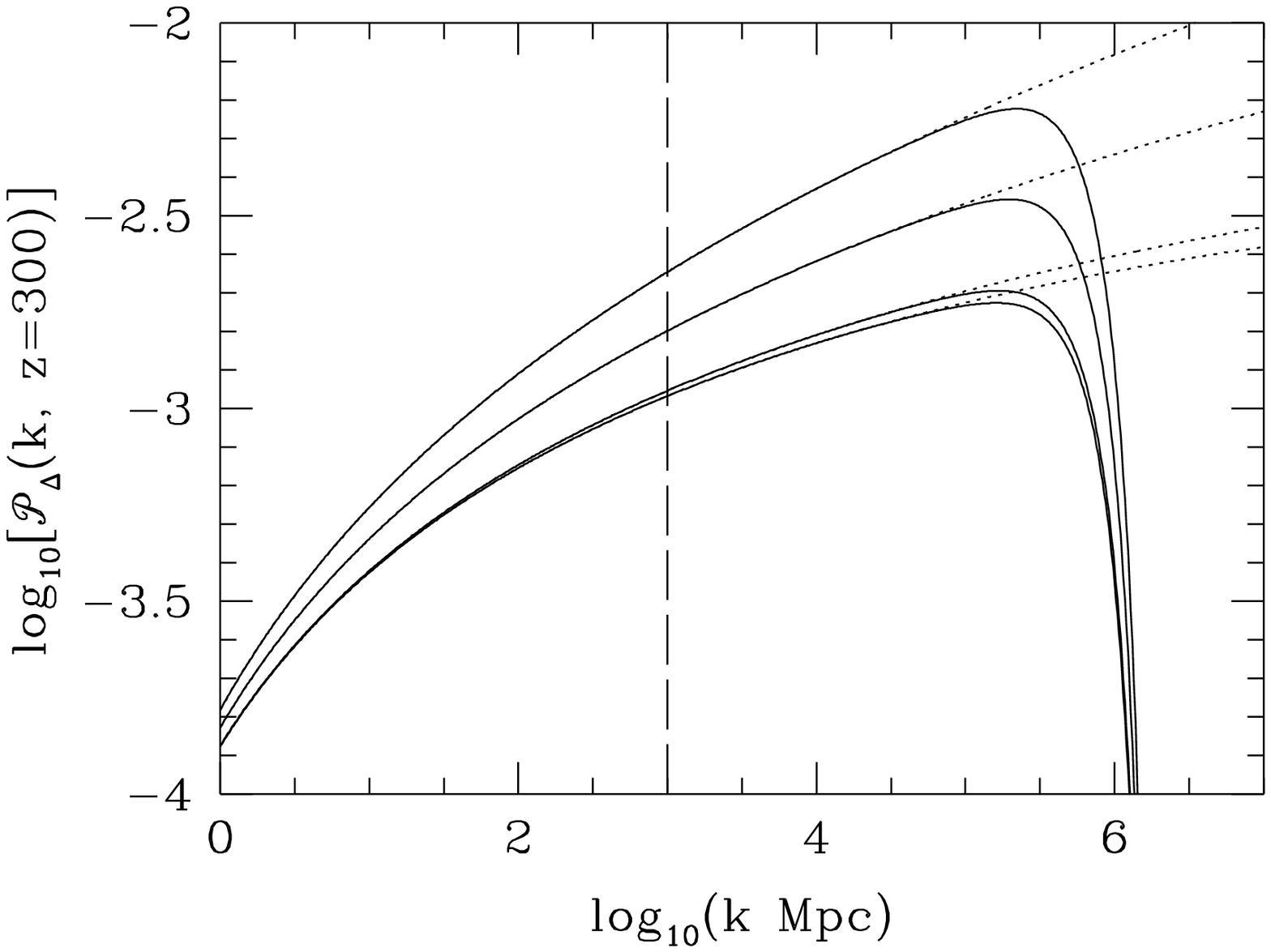, width=5 cm}
\caption{CDM power specrum at $z= 300$ for four different WIMP models (left) and different scenarios of 
cosmological inflation (right) [from Green et al.\cite{GHS2}].  The dotted lines show the power spectra without damping due to CDM microphysics.}
\end{figure}

\subsection{The first objects}

According to theory and consistent with present observations,  primordial fluctuations are gaussian 
distributed.  The growth of matter perturbations on small scales (compared to the Hubble horizon) 
is surpressed during the radiation dominated epoch --- they grow logarithmically. After matter-
radiation equality all subhorizon fluctuations grow proportional to the scale factor and collapse once the overdensity becomes non-linear. For approximately scale-invariant primordial perturbations, these ingredients lead to the hierarchical scenario of structure formation. The smallest scales are the first 
that start to grow and are thus the first to reach the nonlinear regime. The redshift when that happens is estimated for the scales $M > M_{\rm fs}$ to be  $z_{\rm nl} =  40$ to $80$, depending on the details of the WIMP microphysics and the details of the primordial power spectrum. 

The very first objects however are formed much earlier, as they are not typical. For gaussian distributed inhomogeneities the $N\sigma$ fluctuations would collapse at $N z_{\rm nl}$. 
Let us look at some numbers for $N=3$ (see \cite{GHS2}): 
consider the comoving volume the Milky Way
with a mass collecting radius of about $1$ Mpc. Let us fix the scale of the very first objects to be 
$10^{-6} M_\odot$ and let us assume that $z_{\rm nl} = 60$. The density contrast of these CDM clouds  would be $\Delta \sim 10^7$ today, which is an order of magnitude above the mean density contrast of the galactic disc. There would be CDM cloud in a $\mbox{pc}^3$ within the galaxy. However, the
survival of those objects has to be understood before a conclusion for their observational relevance can
be drawn. These studies must include numerical simulations and analytic estimates 
\cite{Diemand,BDE,Green,Goerdt}.  

In a numerical simulation by Diemand et al. \cite{Diemand}, it was shown that indeed structure on all 
scales down to the cut-off scale survive the merging of the first halos. The typical fluctuations of  
a given mass scale merge to give larger ones, however the rare ($N\sigma$, $N > 1$) fluctuations are 
denser and are less likely to be disrupted in the merger process. Rare fluctuations 
survive with a certain probability until the formation of the first  stars. Once compact baryonic objects (stars and planets) have been formed, encounters of them with 
the smallest CDM clouds could lead to their disruption\cite{Silk}. 
This process was studied in some detail in 
references \cite{Green, Goerdt}, and the conclusion is that the clouds would loose mass, but their core 
remains intact, i.e. the cores of the very first objects in the Universe might populate our and other 
galaxies today. Their distribution in mass, size and space is still a subject of research.    
  
\section{Non-thermal and non-relativistic dark matter}

The best studied candidate for non-thermal CDM is probably the axion 
\cite{Sikivie}. Let us just mention that in the case of axions there is a 
preferred mass scale on which miniclusters of $\sim 10^{-12} M_\odot$ are 
formed \cite{Hogan}. 

\section{Effects from the hot plasma}

A sequence of transitions, like the QCD transition, took place in the radiation dominated epoch of the 
Universe \cite{1stsec}. Especially the QCD transition \cite{SSW1} and the $e^\pm$ annihilation 
\cite{Bertschinger} are of interest to the structure formation at the smallest scales. Extra features are imprinted in the CDM power spectra at scales of $10^{-8} M_\odot$ (most relevant in the for non-thermal CDM) and $50 M_\odot$ (relevant for any CDM), respectively.   

\section*{Acknowledgements}
I would like to thank Anne Green and Stefan Hofmann for our enjoyable collaboration and  Ed 
Bertschinger  for a very useful e-mail conversation.

\end{document}